\documentclass[aps,prc,twocolumn,groupedaddress]{revtex4}

\usepackage{graphicx}
\usepackage{longtable}
\def\vc#1{\mbox{\boldmath $#1$}}

\begin{document}


\title{Analysis of previous microscopic calculations for second $0^+$ state 
in $^{12}$C in terms of 3-alpha particle Bose-condensed state}


\author{Y.~Funaki$^1$, A.~Tohsaki$^2$, H.~Horiuchi$^1$, P.~Schuck$^3$, 
and G.~R\"opke$^4$}
\affiliation{$^1$ Department of Physics, Kyoto University, Kyoto 606-8502, Japan}
\affiliation{$^2$ Department of Fine Materials Engineering, Shinshu University, 
       Ueda 386-8567, Japan}
\affiliation{$^3$ Institut de Physique Nucl\'eaire, 91406 Orsay Cedex, France}
\affiliation{$^4$ FB Physik, Universit\"at Rostock, D-18051 Rostock, Germany}

\date{\today}

\begin{abstract}
 The wave function of the second $0^+$ state of $^{12}$C which was 
 obtained long time ago by solving the microscopic 3$\alpha$ problem is 
 shown to be almost completely equivalent to the wave function of the 
 3$\alpha$ condensed state which has been proposed recently by the present 
 authors. This equivalence of the wave functions is shown to hold in two 
 cases where different effective two-nucleon forces are adopted. This 
 finding gives strong support for interpreting the second  $0^+$ state 
 of $^{12}$C  which is the key state for the synthesis of $^{12}$C 
 in stars ( 'Hoyle' state ), and which is one of the typical mysterious 
 $0^+$ states in light nuclei, as a gas-like structure of three $\alpha$ 
 particles, Bose-condensed into an identical s-wave function.

\end{abstract}



\maketitle

The $\alpha$ clustering nature of the nucleus $^{12}$C has been studied by 
many authors using various approaches\ \cite{carbon}. Among these studies, solving the fully microscopic three-body problem of $\alpha$ 
clusters gives us the most important and reliable theoretical information of 
$\alpha$ clustering in $^{12}$C within the assumption that no $\alpha$ 
cluster is distorted or broken except for the change of the size parameter 
of the $\alpha$ cluster's internal wave function. As representatives for the 
solution of the microscopic 3$\alpha$ problem where the antisymmetrization 
of nucleons is exactly treated, we here quote two works: one by Uegaki et al.
\ \cite{uegaki} and the other by Kamimura et al.\ \cite{kamimura} both of 
which were published almost a quarter century ago. In these works, the $^{12}$C 
levels are described by the wave function of the form 
${\cal A} \{\chi({\vc s},{\vc t}) \phi_\alpha^3\}$ with ${\cal A}$ 
standing for the antisymmetrizer, $\phi_\alpha^3 \equiv \phi(\alpha_1) 
\phi(\alpha_2) \phi(\alpha_3)$ for the product of the internal wave 
functions of 3 $\alpha$ clusters, and ${\vc s}$ and ${\vc t}$ for the 
Jacobi coordinates of the center-of-mass motion of 3 $\alpha$ clusters. 
Here $\phi(\alpha_i)$ ($i=1,2,3$) is the internal wave function of the $\alpha$-cluster $\alpha_i$ having the form $\phi(\alpha_i) \propto \exp [ -(1/8b^2) \sum_{m>n}^4 ({\vc r}_{im} - {\vc r}_{in})^2 ]$. 
The wave function $\chi({\vc s},{\vc t})$ of the relative motion of 
3 $\alpha$ clusters is obtained by solving the eigen-energy problem of 
the full three-body equation of motion; $\langle \phi_\alpha^3 | (H-E) | 
{\cal A} \{\chi({\vc s},{\vc t}) \phi_\alpha^3\} \rangle = 0$, where 
$H$ is the microscopic Hamiltonian consisting of the kinetic energy, 
effective two-nucleon potential, and the Coulomb potential between 
protons. The difference between the works by Uegaki et al. and Kamimura 
et al. lies in the adopted effective two-nucleon forces, besides 
the differing techniques of solution.

Both calculations by Uegaki et al. and Kamimura et al. reproduced reasonably 
well the observed binding energy and r.m.s. radius of the ground $0^+_1$ 
state which is the state with normal density, while they both predicted 
a very large r.m.s. radius for the second $0^+_2$ state which is larger 
than the r.m.s. radius of the ground $0^+_1$ state by about 1 fm, i.e. by over 30\%. 
The observed $0^+_2$ state lies slightly above the 3$\alpha$ breakup 
threshold and the energies of the calculated $0^+_2$ state reproduced 
reasonably well the observed value although the value by Uegaki et al. 
is slightly higher than the 3$\alpha$ breakup threshold by 
about 1 MeV.  The second $0^+$ state of $^{12}$C is well known 
as the key state for the synthesis of $^{12}$C in stars ( Hoyle state ) 
and also as one of the typical mysterious $0^+$ states in light nuclei 
which are very difficult to understand from the point of view of the 
shell model. For the understanding of the nature of the $0^+_2$ state 
with dilute density, the analysis by 
Uegaki et al. of the Reduced Width Amplitude (RWA) function of the 
$^8$Be-$\alpha$ breakup is very useful. The RWA function $y_L(\rho)$ 
which is defined as $y_L(\rho) = \sqrt{12 !/ 8 ! 4 !} \langle 
[\Phi({}^8{\rm Be}, L) \phi(\alpha) Y_L({\hat \rho})]_{J=0} | {\cal A} 
\{\chi({\vc s},{\vc t}) \phi_\alpha^3\} \rangle$ with ${\bf \rho}$ standing 
for the realtive coordinate between $^8$Be and $\alpha$, proved to have 
similar magnitude for all partial waves $L$ ( $L$= 0, 2, 4) for the ground 
$0^+_1$ state but it turned out to be large only for $L$=0 for the $0^+_2$ 
state.  This result for the $0^+_2$ state with dilute density implies that 
the $0^+_2$ state has a gas-like structure of 3 $\alpha$-particles 
which interact weakly among one another, predominantly in relative S waves.  
This understanding of the $0^+_2$ state structure had been already presented 
by Horiuchi on the basis of the 3$\alpha$ OCM ( orthogonality condition 
model ) calculation\ \cite{hori}, and is quite different from the picture 
of a 3$\alpha$ linear-chain structure\ \cite{mori} for this state.   
It should be mentioned here that both calculations by Uegaki et al. 
and Kamimura et al. reproduced well not only the energy but also other 
observed quantities related to the $0^+_2$ state indicating that 
their wave fuctions of the $0^+_2$ state are highly reliable.  For example, 
the reduced $\alpha$-decay widths of the $0^+_2$ 
state calculated by Uegaki et al. and Kamimura et al. at the channel radius 
$a$ = 7 fm are 0.39 and 0.56, respectively, while the observed value is 
0.38. The calculated values of the monopole matrix element 
$M(0^+_2 \rightarrow 0^+_1)$ by Uegaki et al. and Kamimura et al are 
6.6 fm$^2$ and 6.7 fm$^2$, respectively, while the observed value is 
5.4 fm$^2$.

Recently, based on the investigations by R\"opke, Schuck, and coauthors on the 
possibility of $\alpha$-particle condensation in low-density nuclear 
matter\ \cite{roepke}, the present authors 
proposed a conjecture that near the $n\alpha$ threshold in self-conjugate 
$4n$ nuclei there exist excited states of dilute density which are composed 
of a weekly interacting gas of self-bound $\alpha$ particles and which can be considered 
as an $n\alpha$ condensed state\ \cite{thsr}. This conjecture was backed by 
examining the structure of $^{12}$C and $^{16}$O using a new 
$\alpha$-cluster wave function of the $\alpha$-cluster condensate type.
The new $\alpha$-cluster wave function actually succeeded to place a level 
of dilute density ( about one third of ground state density ) in each system 
of $^{12}$C and $^{16}$O in the vicinity of the 3 respectively 4 $\alpha$ 
breakup threshold, without using any adjustable parameter.  In the case 
of $^{12}$C, this success of the new $\alpha$-cluster wave function may 
seem rather natural because, as we explained above, we had already known 
that the microscopic 3$\alpha$ cluster models had predicted that 
the $0^+_2$ in the vicinity of the 3$\alpha$ breakup threshold has a 
gas-like structure of $3\alpha$-particles which interact weakly with 
each other predominantly in relative $S$ waves.

The new $\alpha$-cluster wave function of the $\alpha$-cluster condensate 
type used in Ref.\ \cite{thsr} represents a condensation of $\alpha$-clusters 
in a spherically symmetric state. The present authors extended the wave 
function so that it can describe the $\alpha$-cluster condensate with spatial 
deformation\ \cite{fhtsr}. They applied this new wave function to $^8$Be 
and succeeded to reproduce not only the binding energy of the ground 
state but also the energy of the excited $2^+$ state. In addition, 
they found that although the effect of the spatial deformation is not 
large, the introduction of the spatial deformation brought forth a 
100 \% overlap of the condensate wave function with the ''exact'' wave 
function given by the microscopic 2$\alpha$ cluster model which solves 
the 2$\alpha$-cluster equation of motion, 
$\langle \phi_\alpha^2 | ( H - E ) | {\cal A} \{ \chi ( {\vc r} ) 
\phi_\alpha^2 \} \rangle = 0$.  This fact forces us to modify our 
understanding of the $^8$Be structure from the 2$\alpha$ 'dumb-bell' 
structure to the 2$\alpha$ dilute ( gas-like ) structure.

The purpose of this short note is to report on our study of $^{12}$C  
using the extended 3$\alpha$ condensate wave function with 
spatial deformation and comparing the obtained results for $^{12}$C 
with those of the ''exact'' 3$\alpha$ cluster model wave functions 
by Uegaki et al. and by Kamimura et al.  The most remarkable result of 
this comparison is that the $0^+_2$ wave functions by Uegaki et al. and 
by Kamimura et al. are almost completely equivalent to our condensate wave 
functions with slight spatial deformation which are obtained 
by using the same effective two-nucleon force as Uegaki et al. and Kamimura 
et al., respectively. This result implies that the ''exact'' 3$\alpha$ cluster 
model wave functions for the second $0^+_2$ state of $^{12}$C can definitely 
be interpreted as 3$\alpha$-particle Bose-condensed state.

The wave function of the $n \alpha$-cluster condensate with spatial 
deformation was introduced in Ref.\ \cite{fhtsr} and the detailed 
explanation of it is given there. So here we give a brief explanation 
which is necessary in this paper. The wave function has the form 
\begin{eqnarray*}
\lefteqn{
 \Phi_{n\alpha}(\beta_x, \beta_y, \beta_z) =  \int \! d^3R_1 \cdots d^3R_n } \\ 
 &&\!\!\!\!\times \exp \Big\{ -\sum_{i=1}^n \Big( \frac{R_{ix}^2}
 {\beta_x^2} + \frac{R_{iy}^2}{\beta_y^2} + \frac{R_{iz}^2}{\beta_z^2} \Big) 
 \Big\} \Phi^{\rm B} ({\vc R}_1, \cdots, {\vc R}_n) \\
 \propto &&\!\!\!\!\!\!{\cal A} \Big[ \exp \Big\{\!\!-\!\!\sum_{i=1}^n \Big( \frac{2X_{ix}^2}
 {B_x^2}\!+\!\frac{2X_{iy}^2}{B_y^2}\!+\!\frac{2X_{iz}^2}{B_z^2} \Big) \Big\}\phi(\alpha_1) \cdots \phi(\alpha_n) \Big], 
\end{eqnarray*}
where ${\vc X}_i = (1/4) \sum_{n=1}^4 {\vc r}_{in}$ is the center-of-mass 
coordinate of the $i$th $\alpha$-cluster $\alpha_i$, $\phi(\alpha_i)$ is the same internal wave function of the $\alpha$-cluster $\alpha_i$ as the previous microscopic $3\alpha$ cluster model, $B_k^2=b^2 + 2\beta_k^2$ \ ($k= x, y, z$), and 
$\Phi^{\rm B} ({\vc R}_1, \cdots, {\vc R}_n)$ is Brink's $\alpha$-cluster 
model wave function\ \cite{brink}.  It is to be noted that 
$\Phi_{n\alpha}(\beta_x, \beta_y, \beta_z)$ expresses the state where $n$ 
$\alpha$-clusters occupy the same spatially deformed center-of-mass 
orbit $\exp [ -(2/B_x^2)X_x^2 - (2/B_y^2)X_y^2 - (2/B_z^2)X_z^2 ]$, 
while the internal $\alpha$-cluster wave functions stay spherical. 
$\Phi_{n\alpha}(\beta_x, \beta_y, \beta_z)$ can be written as a product of 
the total center-of-mass wave function and the internal wave function 
${\widehat \Phi}_{n\alpha}(\beta_x, \beta_y, \beta_z)$ as 
\begin{eqnarray*}
\lefteqn{
 \Phi_{n\alpha}(\beta_x, \beta_y, \beta_z) \propto
 \exp \Big\{\! -\!\frac{2nX_{Gx}^2}{B_x^2}\! -\! \frac{2nX_{Gy}^2}{B_y^2}\! 
 -\! \frac{2nX_{Gz}^2}{B_z^2} \Big\} } \\
 &&\hspace{3cm}\times{\widehat \Phi}_{n\alpha}(\beta_x, \beta_y, \beta_z), \\
\lefteqn{
 {\widehat \Phi}_{n\alpha}(\beta_x, \beta_y, \beta_z) = {\cal A} \Big[ \exp 
  \Big\{\!-\!\sum_{i=1}^n \Big( \frac{2}{B_x^2} (X_{ix}\! -\! X_{Gx})^2 }\\ 
+\!\!\!&&\!\!\!\frac{2}{B_y^2} (X_{iy}\! -\! X_{Gy})^2\! +\! \frac{2}{B_z^2} (X_{iz}\! -\! X_{Gz})^2  \Big) \Big\} \phi(\alpha_1) \cdots \phi(\alpha_n) \Big].
\end{eqnarray*}
All the calculations are made not with
$\Phi_{n\alpha}(\beta_x, \beta_y, \beta_z)$ but with
${\widehat \Phi}_{n\alpha}(\beta_x, \beta_y, \beta_z)$ which is an eigen state of total momentum with eigen value zero.
In this paper we assume axial symmetry of the deformation around 
the intrinsic $z$-axis and put $\beta_x = \beta_y$.  The $\alpha$-condensed 
wave function with good angular momentum which is obtained by spin projection 
is then written as 
$$
 {\widehat \Phi}_{n\alpha}^J(\beta_x\!=\!\beta_y, \beta_z)\! = \!\!\int \!\! 
 d\cos \theta d_{M 0}^J(\theta) {\widehat R}_y (\theta) 
 {\widehat \Phi}_{n\alpha}(\beta_x\!=\!\beta_y, \beta_z),
$$
where ${\widehat R}_y (\theta)$ is the rotation operator around the 
intrinsic $y$ axis which rotates ${\widehat \Phi}_{n\alpha}$ by an angle 
$\theta$, and $d_{M 0}^J(\theta)$ is the small $D$-function. 

As effective two-nucleon forces, we adopt the ones of Uegaki et 
al. and Kamimura et al.  One is the Volkov force No.1\ \cite{volkov} 
with Majorana parameter $M=0.575$, used by Uegaki et al., and the 
other is the Volkov force No.2\ \cite{volkov} with Majorana parameter 
$M=0.59$, used by Kamimura et al.
Hereafter, we refer to the former force as force I while the latter force 
is referred to as force II.  We adopt the same values for the 
oscillator parameter $b$ as Uegaki et al. and Kamimura et al., namely 
$b=1.41$ fm for force I while $b=1.35$ fm for force II.

In Fig.1 we give the contour map of the $J^\pi = 0^+$ binding energy 
surface corresponding to the spin-projected state 
${\widehat \Phi}_{3\alpha}^{J=0}(\beta_x=\beta_y, \beta_z)$ in the two 
parameter space, $\beta_x (=\beta_y)$ and $\beta_z$.  The adopted effective 
force for this energy surface is force II.   We see the energy minimum 
at $\beta_x (=\beta_y)$ = 1.5 fm and $\beta_z$ = 1.5 fm, which means that 
the minimum has a spherical shape.  The minimum 
energy of $-87.68$ MeV is about 1.7 MeV higher than the binding energy 
of $-89.4$ MeV obtained by Kamimura et al. for the ground $0^+_1$ state. 
The energy surface in the case of force I is similar to the energy 
surface of Fig.1.  The minimum energy obtained by the use of force I 
is $-86.09$ MeV and it is about 1.8 MeV higher than the binding energy 
of $-87.92$ MeV obtained by Uegaki et al. for the ground $0^+_1$ state.

\begin{figure}
\includegraphics{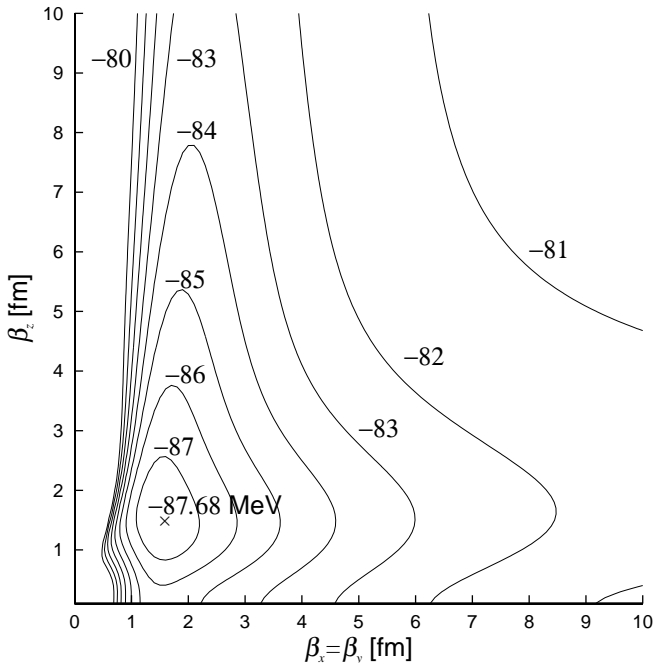}
\caption{Contour map of the energy surface of the $0^+$ state 
in the two parameter space, $\beta_x (=\beta_y)$ and $\beta_z$. 
The adopted effective force is force II.\label{Fig1}}
\end{figure}

\begin{figure}
\includegraphics{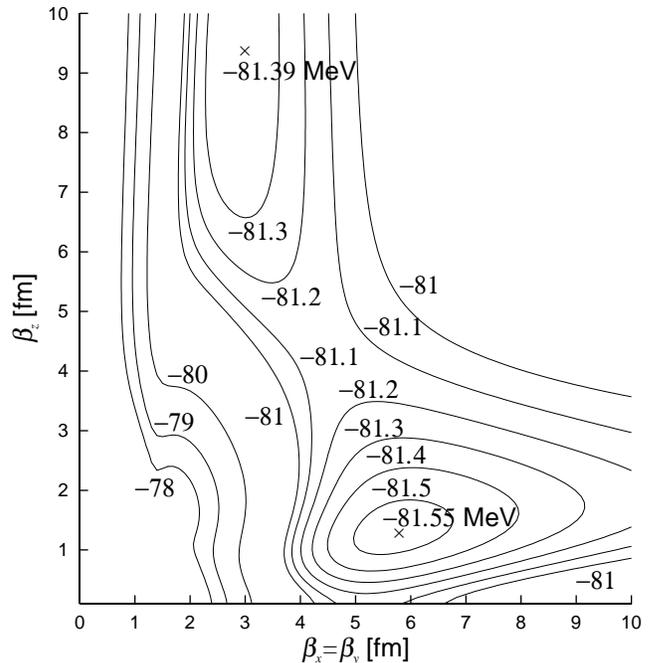}
\caption{Contour map of the energy surface corresponding to the $0^+$ 
state orthogonalized to the state at the minimum energy point in 
Fig. 1. The adopted effective force is force II.\label{Fig2}}
\end{figure}

In Fig.2 we give the contour map of the $J^\pi = 0^+$ binding energy surface 
corresponding to the state orthogonalized to the minimum energy state 
${\widehat \Phi}_{3\alpha}^{J=0}({\rm min.})$ ( the state at the 
minimum energy point in Fig.1). The orthogonalized state is denoted as 
$P_\perp {\widehat \Phi}_{3\alpha}^{J=0}(\beta_x=\beta_y, \beta_z)$ and is 
expressed as 
\begin{eqnarray*}
\lefteqn{P_\perp {\widehat \Phi}_{3\alpha}^{J=0}(\beta_x\!=\!\beta_y, \beta_z)}\\ 
&&\!\!\!\!\!\! =\!( 1\! -\! |{\widehat \Phi}_{3\alpha}^{N,J=0}({\rm min.}) \rangle 
 \langle {\widehat \Phi}_{3\alpha}^{N,J=0}({\rm min.}) | )
 {\widehat \Phi}_{3\alpha}^{J=0}(\beta_x\!=\!\beta_y, \beta_z), \\
\lefteqn{{\widehat \Phi}_{3\alpha}^{N,J=0}({\rm min.}) \equiv 
 {\widehat \Phi}_{3\alpha}^{J=0}({\rm min.}) / \sqrt{ \langle 
 {\widehat \Phi}_{3\alpha}^{J=0}({\rm min.}) | 
 {\widehat \Phi}_{3\alpha}^{J=0}({\rm min.}) \rangle }}.
\end{eqnarray*} 
The adopted effective force for Fig.2 is force II.  We see an energy 
minimum at $\beta_x (=\beta_y)$ = 5.7 fm and $\beta_z$ = 1.3 fm in the 
oblate region of the map and a second energy minimum at $\beta_x 
(=\beta_y)$ = 2.9 fm and $\beta_z$ = 9.4 fm in the prolate region of the map.  
The minimum energy value is $-81.55$ MeV and, what is very remarkable,  
this value is almost the same as the binding energy of $-81.66$ MeV 
obtained by Kamimura et al. for the second $0^+_2$ state.  The minimum 
energy of $-81.55$ MeV is close to the second minimum energy of $-81.39$ MeV,  
and there is a valley with an almost flat bottom connecting these two 
minima. An almost flat bottom of the valley means that 
the energy of the spherical configuration is only slightly higher than that of the deformed configuration, namely the energy gain due to the deformation is small.  The energy surface by the orthogonalized state 
$P_\perp {\widehat \Phi}_{3\alpha}^{J=0}(\beta_x=\beta_y, \beta_z)$ 
in the case of the force I is similar to the energy surface of Fig.2, 
and the minimum energy of the orthogonalized state is $-79.83$ MeV.  Here 
again it is very remarkable that this value is almost the same as 
the binding energy of $-79.3$ MeV obtained by Uegaki et al. for the second 
$0^+_2$ state. 

The fact that for each case of the two different effective forces a single 
orthogonalized state $P_\perp {\widehat \Phi}_{3\alpha}^{J=0}(\beta_x=
\beta_y, \beta_z)$ yields almost the same energy as the ''exact'' energy 
of the $0^+_2$ state obtained by solving a full three-body problem of the 
microscopic 3$\alpha$-cluster model, strongly suggests that the $0^+_2$ state 
wave function given by the microscopic 3$\alpha$-cluster model is similar 
to the rather simple state 
$P_\perp {\widehat \Phi}_{3\alpha}^{J=0}(\beta_x=\beta_y, \beta_z)$ 
as long as the adopted effective two-nucleon force is reasonably realistic. 

We also calculated the $J^\pi = 2^+$ energy surface corresponding to the 
spin-projected state 
${\widehat \Phi}_{3\alpha}^{J=2}(\beta_x=\beta_y, \beta_z)$ for 
the forces I and II.  The minimum energies for the forces I and II 
are obtained to be $-83.61$ MeV at $\beta_x (=\beta_y)$ = 1.30 fm and 
$\beta_z$ = 0.35 fm, and to be $-84.65$ MeV at $\beta_x (=\beta_y)$ = 1.50 fm 
and $\beta_z$ = 0.35 fm, respectively. These minimum energy values for forces I and II are both higher by about 2 MeV than the lowest 
$2^+$ energies by Uegaki et al. and by Kamimura et al., respectively, 
whose values are shown in Table I.

\begin{table*}
\begin{ruledtabular}
\begin{center}
\caption{Comparison of the minimum energy of the spin-projected energy 
surface, GCM eigen energy, and the energy given by the full $3\alpha$ 
calculation. Comparison is made for the $0^+_1$, $0^+_2$, and $2^+_1$ 
states for two cases of the effective two-nucleon force.  The energy 
surface for the $0^+_2$ state means that of the orthogonalized state 
$P_\perp{\widehat \Phi}_{3\alpha}^{J=0}(\beta_x=\beta_y, \beta_z)$. 
Energies are in MeV. }
\begin{tabular}{ c c c c c c c } 
 & \multicolumn{3}{ c }{{\rm Volkov No.1} $M=0.575,\ E_{\rm th}(3\alpha)=-81.01$}
 & \multicolumn{3}{ c }{{\rm Volkov No.2} $M=0.59 ,\ E_{\rm th}(3\alpha)=-82.04$} \\
 \hline
 & $E_{\rm min}$ of & GCM & full $3\alpha$
 & $E_{\rm min}$ of & GCM & full $3\alpha$ \\
 & energy surface & eigen energy & calculation\ \cite{uegaki} 
 & energy surface & eigen energy & calculation\ \cite{kamimura} \\
 \hline
 $0^+_1$ & $-86.09$ & $-87.81$ & $-87.92$ & $-87.68$ & $-89.52$ & $-89.4$ \\
 $0^+_2$ & $-79.83$ & $-79.97$ & $-79.3$ & $-81.55$ & $-81.79$ & $-81.7$ \\
 $2^+_1$ & $-83.61$ & $-85.34$ & $-85.7$ & $-84.65$ & $-86.71$ & $-86.7$ \\
\end{tabular}
\end{center}
\label{tab1}
\end{ruledtabular}
\end{table*}

We also performed the GCM ( generator coordinate method ) calculation for 
$J^\pi = 0^+$ and $2^+$ by superposing ${\widehat \Phi}_{3\alpha}^J
(\beta_x=\beta_y, \beta_z)$ over various sets of $(\beta_x, \beta_z)$; 
\begin{eqnarray*}
\sum_{(\beta_x, \beta_z)} \langle {\widehat \Phi}_{3\alpha}^J
(\beta_x'=\beta_y', \beta_z') | ( H - E_k) | {\widehat \Phi}_{3\alpha}^J
(\beta_x=\beta_y, \beta_z) \rangle && \\ 
\times f^J_k (\beta_x, \beta_z) = 0.&&
\end{eqnarray*}
The adopted values of $\beta_x$ are $\beta_x = (i - 0.5 )$ fm with $i = 
1 \sim 6$, and those of $\beta_z$ is $\beta_z = (j - 0.5 )$ fm with $j = 
1 \sim 8$. Hence the total number of the adopted grid points 
$(\beta_x, \beta_z)$ is 48.  The calculated eigen energies of the 
$0^+_1$, $0^+_2$, and $2^+_1$ states are given in Table I for the two 
forces I and II.   We have checked the 
convergence of the calculation of the eigen energies by changing the sets of 
$(\beta_x, \beta_z)$ for the GCM calculation. 
We see in Table I that all the GCM eigen energies of the $0^+_1$, $0^+_2$, 
and $2^+_1$ states are almost the same as the energies of the microscopic 
3$\alpha$-cluster model in both cases of forces I and II.  Since the 
eigen energies obtained by solving the full three-body problem of the 
microscopic 3$\alpha$-cluster model are the "exact" energies, we can say 
by using the mini-max theorem of the variational problem that this almost 
complete equivalence of our GCM energies with the exact energies means that 
our GCM wave functions of the $0^+_1$, $0^+_2$, and $2^+_1$ states are 
almost equivalent respectively to the $0^+_1$, $0^+_2$, and $2^+_1$ wave 
functions of the microscopic 3-$\alpha$-cluster model in both cases of 
force I and II.  In order to check further this almost complete equivalence, 
we give in Table II the comparison of the calculated 
r.m.s. radii and monopole matrix elements $M(0^+_2 \rightarrow 0^+_1)$ 
between our GCM and the microscopic 3$\alpha$-cluster model. We see nice 
agreement of the calculated quantities between our GCM and the microscopic 
3$\alpha$-cluster model. In Table II, we see that the large r.m.s. radius of 
the $0^+_2$ state is also predicted by our GCM as by the microscopic 
3$\alpha$-cluster model, but at the same time we see that the calculated 
value corresponding to our GCM is slightly larger than that of the 3$\alpha$-cluster model. We think the reason is because our wave function of the $0^+_2$ state 
which contains a large amount of the components of the 3$\alpha$ condensed wave 
functions ${\widehat \Phi}_{3\alpha}^{J=0}(\beta_x=\beta_y, \beta_z)$ with 
large $\beta_x$ and/or $\beta_z$  so as to yield large r.m.s. radius may have 
a longer tail than the former 3$\alpha$-cluster model.  This possibly longer 
tail behavior of the GCM $0^+_2$ wave function may explain the slight 
underestimation of the monopole matrix element of the GCM versus the 
3$\alpha$-cluster model through the slightly enhanced mismatch between 
the $0^+_1$ and $0^+_2$ wave functions in the GCM case.

\begin{table*}
\begin{ruledtabular}
\begin{center}
\caption{Comparison of the r.m.s. radii $R_{\rm rms}$ and the monopole 
matrix element $M(0^+_2 \rightarrow 0^+_1)$ obtained by the GCM 
calculation with those by the full $3\alpha$ calculation. 
Comparison is made for two cases of the effective two-nucleon force. 
$R_{\rm rms}$ are in fm, and $M(0^+_2 \rightarrow 0^+_1)$ are in fm$^2$.}



\begin{tabular}{c c c c c } 
 & \multicolumn{2}{ c }{{\rm Volkov No.1} $M=0.575$}
 & \multicolumn{2}{ c }{{\rm Volkov No.2} $M=0.59$} \\
  \hline
 & GCM calculation& full $3\alpha$ calculation\ \cite{uegaki}
 & GCM calculation& full $3\alpha$ calculation\ \cite{kamimura} \\
 \hline
 $R_{\rm rms}(0^+_1)$ & 2.40 & 2.53 & 2.40 & 2.40 \\
 $R_{\rm rms}(0^+_2)$ & 4.44 & 3.50 & 3.83 & 3.47 \\
 $R_{\rm rms}(2^+_1)$ & 2.38 & 2.50 & 2.38 & 2.38 \\
 $M(0^+_2 \rightarrow 0^+_1)$ & 5.36 & 6.6 & 6.45 & 6.7 \\
\end{tabular}
\end{center}
\label{tab2}
\end{ruledtabular}
\end{table*}

The fact that the second $0^+_2$ wave function of the microscopic 
3$\alpha$-cluster model is almost completely equivalent to our GCM wave 
function of the second $0^+_2$ state which has a very large r.m.s. radius 
or equivalently very dilute density is very important. Since our GCM wave 
function of the $0^+_2$ state expresses the Bose-condensed state of 
3$\alpha$-clusters, as is clear from its large r.m.s. radius and from its 
functional form, we can say that the second $0^+_2$ wave function of 
the microscopic 3$\alpha$-cluster model obtained long time ago underlines 
the fact that the second $0^+$ state of $^{12}$C in the vicinity of the 
3$\alpha$ breakup threshold has a gas-like structure of 3$\alpha$ clusters 
with ''Bose-condensation''.

Now we discuss the relation between our GCM wave function of the $0^+_2$ 
state which we denote as $\Psi_{\rm GCM}(0^+_2)$ and the orthogonalized state  
$P_\perp {\widehat \Phi}_{3\alpha}^{J=0}(\beta_x=\beta_y, \beta_z)$ 
with minimum energy which we denote by $\Psi_\perp(0^+_2)$. Although the 
energy of $\Psi_\perp(0^+_2)$ is almost equivalent to that of 
$\Psi_{\rm GCM}(0^+_2)$ and also to that of the $0^+_2$ wave function of 
the microscopic 3$\alpha$ cluster model, 
we cannot simply conclude that $\Psi_\perp(0^+_2)$ 
is almost equivalent to $\Psi_{\rm GCM}(0^+_2)$.  It is because 
$\Psi_\perp(0^+_2)$ is not yet guaranteed to be orthogonal to the 
$0^+_1$ wave function.  The orthogonality of $\Psi_\perp(0^+_2)$ to 
${\widehat \Phi}_{3\alpha}^{J=0}({\rm min.})$ which is the state at the 
minimum energy point of the energy surface is not the same as the 
orthogonality to the $0^+_1$ wave function, and $\Psi_\perp(0^+_2)$ 
may contain some amount of the $0^+_1$ wave function.  
We therefore calculated the squared overlap value of the two wave functions, 
$|\langle \Psi_\perp(0^+_2) | \Psi_{\rm GCM}(0^+_2) \rangle |^2$.
The obtained values are 0.95 and 0.97 for 
forces I and II, respectively. These large overlap values mean that the 
GCM $0^+_2$ wave functions are very similar to $\Psi_\perp(0^+_2)$ in both 
cases of force I and II and hence verify our former statement that the 
$0^+_2$ wave function of the microscopic $3\alpha$ cluster model is very 
similar to a simple state $\Psi_\perp(0^+_2)$ so long as the adopted two-
nucleon force reasonably describes the physics.

We also studied the magnitude of the spherical condensate component contained in 
our GCM $0^+_2$ wave functions. For this purpose, we first constructed 
the projection operator $P_{\rm sph}$ onto the functional space S$_{\rm sph}$ 
spanned by spherical condensate wave functions as $P_{\rm sph} = \sum_k 
|\Psi^k_{\rm sph} \rangle \langle \Psi^k_{\rm sph} |$, where 
$\Psi^k_{\rm sph}$ are the orthonormal basis functions of the space 
S$_{\rm sph}$. $\Psi^k_{\rm sph}$ are constructed as follows,
\begin{eqnarray*}
 && \sum_{\beta_x} \langle {\widehat \Phi}_{3\alpha}^{J=0}(\beta_x'=\beta_y'=
 \beta_z') | {\widehat \Phi}_{3\alpha}^{J=0}(\beta_x= \beta_y= \beta_z)
 \rangle g^k(\beta_x) \\ 
 && \hspace{6.3cm}= \mu_k g^k(\beta_x'), \\ 
 && \sum_{\beta_x} g^{k'}(\beta_x) g^k(\beta_x)=\delta_{\beta_{k'} \beta_k}, \\
 && \Psi^k_{\rm sph} =\frac{1}{\sqrt{\mu_k}} \sum_{\beta_x} g^k(\beta_x) 
 {\widehat \Phi}_{3\alpha}^{J=0}(\beta_x= \beta_y= \beta_z).
\end{eqnarray*}
The calculated values of $|\langle \Psi_{\rm GCM}(0^+_2)|P_{\rm sph}|
\Psi_{\rm GCM}(0^+_2)\rangle|^2$ are 0.92 and 0.91 for forces I and II, 
respectively. Of course, we checked the convergence of the calculation by 
changing the number of the adopted components 
$\Psi^k_{\rm sph}$ in $P_{\rm sph}$. The large magnitudes of these 
values imply that $\Psi_{\rm GCM}(0^+_2)$ is 
mostly composed of the spherical condensate component by more than 91 \%. 
At the same time we have to note that some amount (less than 9 \%) of 
the deformed component which is orthogonal to the spherical component   
is necessary in order to have quantitatively good reproduction of the 
observed quantities.

We finally make a remark on the ground $0^+_1$ wave function. Since this 
state has a normal radius and density, three $\alpha$ clusters overlap 
strongly with each other in this state, which is totally different from 
the situation of the $0^+_2$ state where the mutual overlap of three, or 
even two $\alpha$ clusters is small.  Therefore even though the $0^+_1$ state is  well represented by a superposition of our condensate wave functions (we recall that our wave function contains the Slater determinant as a limit case), 
it does not mean at all that the state has an $\alpha$ condensation 
character which is only valid for the gas-like state of $\alpha$-clusters.

In summary, we have shown that the $0^+_2$ wave function of $^{12}$C which 
was obtained long time ago by solving the full three-body problem of the 
microscopic $3\alpha$ cluster model is almost completely equivalent to 
the wave function of the $3\alpha$ condensed state. This equivalence has been  
shown to hold for two different effective two-nucleon forces. 
This result gives us strong support to our opinion that the 
$0^+_2$ state of $^{12}$C has a gas-like structure of $3\alpha$ clusters 
with ''Bose-condensation''. A more detailed report of the present 
problem will be given elsewhere.

\end{document}